# Not Only Horses Wear Blinkers:
# The Missing Perspectives in IS Research

Roger Clarke
Principal, Xamax Consultancy Pty Ltd, Canberra
Visiting Professor in Cyberspace Law & Policy, UNSW Law
Visiting Professor , Research School of Computer Science, Australian National University
Email: Roger.Clarke@xamax.com.au

## Abstract

When we devise a method to address a particular research question, we think about many things, including the unit of study. But do we think about the perspective from which we observe whatever it is that's within our field of view?

In the majority of mainstream IS research, one perspective dominates. Organisations use IS and IT as a means of intervening into a context. And the behaviour of the phenomena is mostly observed from the viewpoint of those organisations.

Is that viewpoint the only legitimate one for IS research to adopt? Even if it is, are the clients' interests best served by IS research that adopts that perspective alone?

This paper identifies alternative perspectives. It draws attention to negative consequences of wearing blinkers to restrict our field of view, and to opportunities we can grasp by taking off the blinkers. It proposes that these alternatives need to be defined as being within the IS discipline rather than outside it.

## 1   Introduction

IS academics spend a great deal of their time on the conception, design, conduct and reporting of research. The many aspects that we invest energy into include existing bodies of theory that can be used as the theoretical lens through which to view phenomena, the formulation of research questions, the selection and application of research techniques to answer those questions, and, particularly in scientific research, population definition, selection of sampling frame and sample, and external validity.

This paper concerns itself with one particular notion that is almost invisible in the research process, almost always left implicit. It is contended that leaving this aspect festering in a dark corner is a serious weakness in our discipline, that this has led to a number of highly unfortunate outcomes, and that we urgently need to throw light on the notion, and to address the issues.

Discussions about research frequently refer to the 'unit of study' or 'unit of analysis'. There are many levels of abstraction at which we can observe phenomena. We choose one, which may be, for example, a category of employees or customers, a particular organisational entity, a category of organisational entities, or an industry segment. We then frame our research questions in terms of that particular unit of study, and develop our research design with it in mind.

The invisible aspect of the research process that is of relevance here is related to the notion 'unit of study', but different from it. The term 'perspective' is adopted in this paper.

## 2   The Concept of Perspective

The term 'perspective', as it is used in this paper, refers to the viewpoint from which phenomena are observed. This is distinct from the frequent mentions encountered in IS research literature, such as a scientific, an interpretivist or a critical-theory perspective, a disciplinary perspective, and a theoretical perspective. The perspectives of interest here are those depicted by the tale about blind men describing an elephant differently, depending on which part of it they are holding. The viewpoints this paper has in focus are practical, empirical or real-world in nature, as seen from the standpoints of stakeholders in the process that is under observation.

Recognition that phenomena look different depending on the vantage-point from which they are viewed has been stronger since interpretivism gained a foothold in IS research. For example "... the intent of [interpretive] research [is] to increase understanding of the phenomenon within cultural and contextual situations; where the phenomenon of interest [is] examined in its natural setting and from the perspective of the participants; and where researchers [do] not impose their outsiders' a priori understanding on the situation" (Orlikowski & Baroudi 1991, p.5, emphasis added). The singular 'perspective of the participants' is misleading, and is more appropriately read as the plural 'perspectives of the participants'.

There are a number of differentiators of interpretivism from scientism (Walsham 1995, Clarke 2000, 2001). Those relevant to the present discussion are the express recognition that:

- the researcher has a perspective;

- the researcher's perspective influences the conception of the research, the expression of the research questions, the research design, the analysis of the evidence, and the results; and

- because multiple interpretations are possible of the same phenomena, the adoption of a single perspective creates a considerable risk of reaching inappropriate conclusions.

For simplicity of expression, it will be assumed during the early part of the discussion that relevant viewpoints are those of particular entities, or categories of entity. That assumption will later be relaxed. For the purposes of this paper, the perspective adopted by a researcher can be determined by identifying or inferring the researcher's conception of the beneficiary/ies of the work.

It is useful to revisit the concept of 'unit of study', and to distinguish it from 'perspective'. Unit of study refers to the object of the verb 'observe'. Hence 'a researcher observes phenomena at the level of abstraction called unit of study'. Perspective, on the other hand, is part of the subject of the sentence, as in 'a researcher adopts a perspective in order to observe phenomena at the level of abstraction called unit of study'. (This approach can be further expanded in order to show the relationship with other aspects of research, e.g. by appending 'using a particular theoretical lens to impose order on the expression of observations').

Searches for the term 'perspective' in leading IS journals identifies limited usage in the sense described here. The word is frequently used to refer to the theoretical lens, but seldom to the interests of a stakeholder. Searches for other terms associated with this concept were also largely unsuccessful, in both text-books on research and in journal articles dealing with research process. Remarkably, the notion appears to be largely absent from the debates about the 'core' of the IS discipline stimulated by Benbasat & Zmud (2003) and Walsham (2012).

The perspective adopted has a very significant influence on the entire research undertaking. As we frame our research, we are adopting a particular point of view, our formulation of research questions is strongly influenced by that view, we evaluate alternative designs using criteria shaped by that perspective, and what is included within and what is excluded from the potential outcomes of the research are oriented towards the interests of the chosen beneficiary, and expressed in ways intended to communicate to the chosen audience.

The following section identifies the particular perspective that dominates IS research. Alternative perspectives are then considered, and their legitimacy evaluated. Approaches to research that actively reflect more than a single perspective are discussed. Negative impacts arising from the overly strong commitment to a single perspective are identified, together with opportunities associated with a more inclusive approach. Implications for the discipline and the profession are identified.

## 3 The Dominant Perspective

Casual observation of the content of IS journals and conference proceedings over a long period has given the author the strong impression that the mainstream perspective adopted in IS research is that of the sponsor of the information system.

By 'system-sponsor' is meant the particular organisation that is developing or implementing the system, or causing it to be developed or implemented, or for whose benefit it has been installed. In some cases, however, the perspective adopted may be that of a category of organisations, such as corporations installing an ERP package or adopting a CRM cloud-service, or hospitals implementing an electronic health record system. In joint ventures and collaborative inter-organisational schemes, the system-sponsor may be a collective. Surveys and experiments are conducted in order to understand the impacts of interventions, from the system-sponsor's perspective. A great deal of interpretivist research is performed in order to provide broader understanding of those impacts. In design science, the notion of design is "the purposeful organization of resources to accomplish a goal" (Hevner et al. 2004). It literally 'goes without saying' that the goal is formulated in order to serve the interests of the organisation(s) by whom or for whom the intervention is made, or the design activities are performed.

Business enterprises define objectives solely in terms of the organisation's own interests. This is 'in the DNA', because the nature of the joint stock company is such that directors have a legal obligation to act in the interests of the company. This mode of thinking has migrated beyond the business context, because mission statements and corporate objectives are now the driving force in public sector and non-profit organisations as well.

The system-sponsor may recognise that other entities also have interests. But those interests are seldom treated as objectives, and operate instead as constraints. And, even then, it is normal for only those entities to be considered that are perceived to be sufficiently powerful that they need to be recognised as stakeholders. Legal requirements and softer forms of influence such as industry codes and Ethics Committees also represent constraints on organisations' freedom of action. However, the last few decades have seen significant winding-back of 'regulation' in favour of mere 'governance', and hence these constraints are in many circumstances very weak. 'Self-regulation' meanwhile is not just the softest form, but is also largely an empty promise by all but the few most strongly values-driven organisations. The 'business ethics' and 'corporate social responsibility' notions are excusatory, designed to keep the threat of regulation and enforcement at bay, not to permit the interests of other parties to become part of the organisation's mission and objectives.

The adoption of the system-sponsor's perspective is clearly both legitimate and valuable. Moreover, it is entirely appropriate in the case of Management Information Systems (MIS) in the US tradition, and Wirtschaftsinformatik (WI – Business Informatics) in German-speaking countries. On the other hand, the broader fields of Information Systems (IS) and Information Management (IM) could reasonably be expected to be more diverse, and to encompass research that adopts perspectives other than that of the system-sponsor. It appears, however, that IS and IM have progressively also adopted the norms that are mainstream in both business schools and software engineering schools.

A later section moves beyond the casual observation on which the analysis in this section was based, and considers some empirical evidence. First, however, it is necessary to identify perspectives that could be alternative, competitive or complementary to those of the system-sponsor.

## 4   Alternative Perspectives

The IS discipline emerged in the mid-1960s, particularly in Scandinavia, the UK, the US and Australia (Clarke 2008, Hirschheim & Klein 2012). This followed the application of computing to administrative tasks, which had commenced in the early 1950s in the UK and the USA. The discipline began by adopting and adapting prior work of relevance, from such areas as Organisation & Methods (O&M), and pre-computing Automated Data Processing (ADP). The initial context was computer-based systems, and the initial challenge was to find ways to manufacture software that enabled systems to perform administrative work very quickly, inexpensively and accurately. So it was natural that there be a strong focus on the interests of the system-sponsor.

However, during the following quarter-century, the impact and implications of IS expanded enormously, as computing progressively married communications to produce IT/ICT, and as the scope of applications extended to all organisational functions, and then to inter-organisational systems. By 1990, extra-organisational systems had emerged, in such forms as videotext, bulletin-board systems, ATMs and EFTPOS, which actively involve not only large organisations but also individuals and micro-businesses (Clarke 1992).

A variety of movements have endeavoured to achieve broader horizons for IS research and practice. The scope of 'system' is far broader than computer-based processes alone, and both human-computer interface and human-performed processes have been incorporated within the discipline and profession. Some approaches have emphasised the interests of individual users within the organisation, and proposed a participative approach to the analysis of requirements and the design of systems (e.g. Land & Hirschheim 1983). As systems reached beyond organisational boundaries, organisations other than the system-sponsor were recognised as having a stake in the design and implementation. As individuals outside the organisation became users of networked computing facilities, their actions and their interests also needed to be encompassed within the frame of reference. As computing became a major consumer of electricity, and as cathode-ray tubes mounded up in scrapyards, impacts on the environment came to notice, and needed to be defined as being within the scope of the IS discipline.

In the contemporary context, a range of possible perspectives can be identified. Any discipline must conduct some research that is internally-facing, such as discussions about scope, method, and quality assurance mechanisms. The IS discipline may be seen as having published an excessive volume of papers whose beneficiary and audience is itself. This has been partly because of its eternal existential angst, but also because, in many segments, rigour has been permitted to overtake relevance. Such genres as the design and validation of survey instruments, and reports on minute variations to prior research, are neither accessible to professionals nor of any use to them. The following sub-sections focus on literature in which relevance to IS practice is the primary criterion and rigour is a vital characteristic rather than the primary objective.

A variety of approaches have been adopted to distinguishing categories of subject-matter. The approach adopted here has some similarities to that in Keller & Coulthard (2013), but differs in a couple of important respects. Three dimensions are identified, each with different levels of abstraction within them. The first – which has already been introduced as including the dominant perspective in IS research – is referred to here as the economic dimension. The two further dimensions are the social and the environmental.

### 4.1 The Economic Dimension

Human activity is coordinated and directed by nested layers of institutional arrangements. Phenomena can be observed from the viewpoint of any of those levels of abstraction. In order to distinguish these viewpoints from the alternatives that follow, the dimension is referred to here as economic.

The previous section identified the interests of individual organisations, or of categories of them, as being the dominant frame of reference evident in IS research. It is also feasible to undertake research from the viewpoint of sub-structures within an organisation, such as work-groups, branch offices and subsidiaries. Above the level of the individual organisation, there are strategic partners, industry segments, value-chains, regional economies, nation-states, supra-national regions (such as the EU, NAFTA and ASEAN), and the world economy. Although they represent a minority, some IS research projects adopt such perspectives, particularly those of clusters of strategic partners, and industry value-chains.

### 4.2 The Social Dimension

Each individual, as a person rather than as a labour resource within economically-oriented organisations, has interests that are affected by information systems. In principle, each person has a perspective from which phenomena can be observed, although in practice it is necessary to consider categories or segments of people.

The idea of 'person' needs to be unpacked, however. It is conventional to apply the term 'user'. Research needs to be conducted from the perspectives of the users of installed software, eCommerce and MCommerce services, government services, Internet Banking, games, SaaS, social media, media and entertainment, etc. User segments can also be recognised that have specific needs, such as the physically-handicapped and the socio-economically disadvantaged. Examples of research questions at this level of abstraction include:

> *What are the characteristics of messaging, workflow and calendar services that ensure a fair work-life balance for employees?*
>
> *What contract terms do consumers need from cloud services such as social media and family-tree web-sites?*

In addition to users, there are also many people who are not directly involved in information systems, yet are affected by them. Such role-categories as job-applicants, credit-applicants, tenants and surveillance subjects are usefully referred to as 'usees' in respect of background databases that are consulted by commercial and administrative decision-makers (Clarke 1992). Research needs to be conducted from the perspective of usees, with other actors, including the system-sponsor, identified as stakeholders, and their interests treated as constraints, not objectives. An example of a research question at this level of abstraction is:

> *What threats to tenants' interests arise from commercial databases that gather and publish data about tenants, and what safeguards, mitigation measures and countermeasures are available to tenants to counter those threats?*

At a higher level of abstraction, a range of institutions exists, within which human activities are coordinated for social rather than economic purposes. People cluster into family, household and kinship groups. They also form geographical communities such as villages, precincts and suburbs, and common-interest communities that may have a physical locus and/or take the form of virtual organisations. They may or may not be incorporated, e.g. as a company, association, cooperative or club. They may arise from a commonality of geography, ethnicity, language, religion or perceived interests. Their purpose may be of an entertainment nature, or they may have a productive, service, representative or advocacy function. An example of a research question at this level of abstraction is:

> *What are the key requirements of a peer-to-peer scheme to accumulate and publish information about the service-levels provided by a particular government agency?*

At an even higher level of abstraction, we conceive of societies and populations, in some cases within geo-political boundaries, in other cases crossing national borders. Most broadly, we can conceive of the perspective of 'world society', or humanity as a whole.

Is it appropriate that the kinds of research questions that arise along the social dimension be investigated within the IS research community?

## 4.3 The Environmental Dimension

A conventional normative standpoint is that interventions of all kinds are intended to be for the benefit of the human race. That assumption is being questioned, with propositions that other forms of life, and the planet as a whole, need to be included within the objective function. The IS discipline must at least recognise the existence and importance of that debate, and arguably should participate in it.

Ecologies need to be recognised, again at various levels of abstraction. At a low level, the impact of CRTs in landfill, and of electricity consumption by computing equipment, can only be discussed if concepts such as individual-species-within-context, local bio-communities, regional bio-communities, or the biosphere as a whole, are regarded as a set of interests from whose perspective phenomena can be observed.

All of the examples given in the two preceding sub-sections involve reasonably definable entities, whether associated with a physical person or with some fiction such as a joint stock company that is widely-recognised, and in some cases even assigned legal rights and responsibilities, such that the lack of a physical existence is overlooked by common consent. Reification also arises in the environmental dimension. For example, 'primitive' societies are recognised as having close associations with nature, as in the specifically Australian Aboriginal notion of 'country'. Echoes reverberate within 'advanced' societies, such as the not-quite-yet-archaic notion of 'Mother Nature'. More modern notions of a similar kind are nature and wilderness reserves, national parks and environmental trusts. An example of a research question at this level of abstraction is:

> *What are the key factors to be considered when evaluating the sustainability profiles of alternative designs for information systems?*

Beyond terrestrial ecologies, climate change issues cannot be meaningfully discussed unless that part of the earth's atmosphere is considered in which weather happens – the troposphere – and it is treated as though it were an entity with a point of view. More abstract notions on the same dimension include the planet – or, more poetically, Spaceship Earth. A research question at this level of abstraction, addressed in Clarke (2009), is:

> *How can eCommerce theory be most effectively applied to carbon trading?*

Again, research from these perspectives is essential. If IS is not the appropriate discipline to consider the environmental impacts and implications of information technologies and systems, which discipline(s) do we defer to?

# 5   The Extent of Adoption in IS Research

In Table 1, some relevant points are identified on each of the three dimensions discussed in the previous section. In order to gain some insight into the extent to which these alternative perspectives are currently evident in IS research, this section adopts two approaches. The first sub-section considers a particular form of research question. The second sub-section reports on a pilot assessment of a small sample of IS publications.

*Table 1.   Dimensions and Levels of Abstraction of Alternative Perspectives*

| **Economic Dimension** | **Social Dimension** | **Environmental Dimension** |
|---|---|---|
| World Economy | Humanity | The Planet |
| Supra-National Region (e.g. EU, NAFTA) | | |
| Nation-State | A Society | The Troposphere |
| Regional Economy | | |
| Sector / Value-Chain | A Community | The Biosphere |
| Strategic Partners | | |
| Organisation | A Person | A Localised Ecology |
| Sub-Organisation | | |

## 5.1   A Test Case

In order to provide a concrete test, the following research question was contrived:

> *What are the impacts ... of the withdrawal of the customer option of receiving printed invoices through the post?*

The base case involves replacing the ellipsis "..." with "*on an organisation*". This adopts the conventional perspective of a corporation or small business enterprise, and appears to be a question of a kind that is mainstream in IS research.

Applying even the points on the three dimensions listed in Table 1 above, over a dozen further variants of the research question can be readily generated, each of whose legitimacy as an IS research question needs to be considered. For example:

> *What are the impacts, <u>ON THE REGIONAL ECONOMY TO WHICH THE ORGANISATION CURRENTLY OUTSOURCES INVOICE PRINTING AND MAILING</u>, of the withdrawal of the customer option of receiving printed invoices through the post?*

> *What are the impacts <u>ON PEOPLE WHO HAVE NO INTERNET CONNECTION</u> of the withdrawal of the customer option of receiving printed invoices through the post?*

> *What are the impacts <u>ON FORESTS</u> of the withdrawal of the customer option of receiving printed invoices through the post?*

It is entirely tenable for these questions to be regarded as intellectually uninteresting, or insufficiently rewarding, or not readily researchable using the techniques commonly applied by the researcher or the institution they work within, or simply lower-priority than other competing topics. The question here, however, is whether such research questions in principle fall inside or outside the scope of the IS discipline.

The author's working definition of the scope of the IS discipline has long been "the multi-disciplinary study of the collection, processing and storage of data; of the use of information by individuals and groups, especially within an organisational context; and of the impact, implications and management of artefacts and technologies applied to those activities" (Clarke 1990).  The author contends that all of the research questions in this test case fall within the scope of that definition, and hence are appropriate subjects for IS research.  Other members of the discipline may, however, disagree, or may feel uncomfortable having such broad responsibilities thrust upon them.

## 5.2   A Pilot Empirical Assessment

Reference was made above to the author's strong impression that the mainstream perspective adopted in IS research is that of the sponsor of the information system.  In order to test that inference from casual observation, it is necessary to inspect a sufficient sample of reports of research, and identify the perspective(s) that each adopt(s).

A pilot study has been performed, in order to gain insight into the challenges that the research design needs to overcome.  Two venues were selected:

- the Australasian Journal of Information Systems (AJIS), Vol. 18 (2014)
- the Australasian Conference in Information Systems (ACIS), 2014

There were 38 papers in the AJIS Volume and 179 in the ACIS Proceedings, so a 20% sample of ACIS 36 papers was extracted in a pseudo-random manner.  The perspectives adopted by authors were identified.  It is unusual for perspective to be formally declared by authors, and in most cases it was necessary for it to be inferred, based primarily on any declared research question, and relevant statements in the introduction and implications sections.  A convenient way to present the outcomes of the inspections is by means of a short-form of the research question that the author addressed.

Each paper was classified by the author into one of seven categories.  Table 2 shows the categories, and the aggregate counts.  Appendices 1 and 2 list the research questions in each of the categories of greatest relevance.  In many cases, the research question was explicit, and in most others it was reasonably clear.  In a range of circumstances, it was necessary to read into the paper in order to ensure appropriate categorisation, in particular to identify who the researcher perceived as being the beneficiary of the work.  Some editing of questions has been performed, to ensure readability when lifted out of context, and to shorten unduly long research questions.

*Table 2.    Categorisation of Papers*

- **Discipline-Internal** paper  
  (e.g. on research method, issues within the discipline, teaching case), not relevant to this study
- **Theoretical** paper  
  (which were in all cases oriented to the system-sponsor perspective)
- Empirical paper on the **Economic** dimension:
  - **Humans do <u>not</u> figure prominently** among the phenomena
  - Humans figure prominently among the phenomena and:
    - **The system-sponsor perspective is predominant**
    - **Some other-party perspective is also apparent**
- Empirical paper on the **Social** dimension
- Empirical paper on the **Environmental** dimension

As Table 3 shows, there were substantial differences in the distributions across the categories in the two venues.  The next phase of this study clearly needs to adopt a careful approach to stratification, sampling frames and samples.

Table 3:   Dominance of the System Sponsor Perspective

| Category | Count | %age | Sub-Sample %ages |
|---|---|---|---|
| Discipline-Internal | 3 | 4 | 5, 3 |
| System-Sponsor Dominant: | | | |
| ▪ Theoretical | 13 | 18 | 8, 28 |
| ▪ Unit of Study <u>not</u> Humans | 31 | 42 | 59, 25 |
| ▪ Unit of Study Humans: | | | |
| ▪ Entirely for the System-Sponsor | 12 | 16 | 5, 28 |
| ▪ Mainly for the System-Sponsor | 12 | 16 | 18, 13 |
| Social Dimension | 3 | 4 | 5, 3 |
| Environmental Dimension | 0 | 0 | 0, 0 |

Within that small sample, a very large percentage of papers do indeed adopt the system-sponsor's perspective (75-90%, depending on the definition adopted), with very small percentages in the other categories.  Some classification difficulties were encountered, in particular in relation to a paper on consumer-to-consumer transactions, and the degree of dominance in the 'not solely for the system-sponsor' category.  Two examples in this category were 'What are the education needs of the business analyst?' and 'What role does social media play in an election?'.

## 6   Opportunities

The recognition of the notion of perspective, and of its importance, creates new potentials for the IS discipline.  The first of these is to increase our scope to encompass **multi-perspective research design**, such as comparison and contrast of the interpretations of IS and IT interventions by two or more parties.  These could be, for example, employer and employees, or supplier and customers, or participants in a hub-and-spoke network.  Aspects of the multi-perspective notion are of course evident in interpretivist research generally.  So the foundations already exist to address research questions of such forms as:

> *What are the impacts <u>ON THE ORGANISATION, AND ON CUSTOMERS</u>, of the withdrawal of the option of receiving printed invoices through the post?*

Such approaches are uncommon in IS research, but not unknown.  Years ago, in the context of international trade EDI, Cameron & Clarke (1996) asked:

> *What are the critical success factors for a project management framework for collaborative inter-organisational systems, from the viewpoints of each of the main players?*

A more recent example is Agarwal et al. (2012).  This examined cyber-collective social movements (CSMs) such as the use of social media in the 'Arab Spring' of 2010-12.  After surveying the available research methods literature, the authors developed an analysis based on Individual Perspective, Community Perspective, and Transnational Perspective.  The paper featured no system-sponsor, but rather three levels of abstraction of the social rather than of the economic dimension discussed earlier in this paper.   This is also noteworthy in being one of only a small number of papers located to date in which the term 'perspective' is used in a manner similar to that proposed in the present paper.

Further, the opportunity exists to extend **constructivist research** techniques.  The way has already been shown by **action research**, because is adopts "the idiographic viewpoint [that] any meaningful investigation must consider the frame of reference and underlying social values of the subjects" (Baskerville 1999).  Design science can follow suit.  The kind of research question implied by Guideline 3 of  Hevner et al. (2004) is of the form:

> *What is a feasible and good design [for a particular system]?*

The accumulated understanding of socio-technical thinking (Emery & Trist 1960, Mumford 2000) can be applied, in order to articulate what might be usefully described as **'participative design science'**, resulting in research questions of the form:

> *What is a feasible and effective process for reflecting the perspectives of all parties in the design [of a particular system or category of systems]?*

Both of those opportunities may benefit from consideration of recent discussions in the literature relating to **'phronesis'**. This Aristotelian notion injects an ethical flavour into research design, by requiring deliberation about the appropriate values that need to be applied, which in turn depends on appreciation of the various perspectives involved and accommodation of the various interests involved (Harrison & Zappen 2003, Constantinides et al. 2012).

These approaches are consistent with a shift in emphasis beyond 'pure research', which is undertaken in order to contribute to abstract, theoretical understanding, and 'applied research', which commences with a tool and uses it to intervene within a context. The notion of **instrumentalist research** does not begin with a tool, but with a perceived problem. The approach is goal-oriented, and involves a search for a solution to that problem (Clarke 2001). Beyond merely acknowledging the existence of multiple perspectives, and describing them, the multi-perspective instrumentalist approach treats each of the parties as a beneficiary, rather than orienting the research question, design, conduct and outcomes to primarily serve the needs of just one of the parties. Hence, an indicative research question becomes:

> *How can an organisation <u>MANAGE</u> the negative impacts <u>ON ALL PARTIES</u> of the withdrawal of the option of receiving printed invoices through the post?*

Once again, the ground has been prepared, in this case by **critical theory research**. This is inherently multi-perspective in nature, being directly concerned with conflicts among the interests of the various actors, and with the power-structures that determine the outcomes. It is also inherently instrumentalist, because it is conducted with the express intention of influencing the phenomena that are under observation (Klein & Myers 1999, McGrath 2005).

The combination of these threads leads to a further and substantial opportunity for the IS discipline. Research from social perspectives has to date only achieved a toehold within the IS literature, and research from the environmental perspective is a rarity. We have to date failed to contribute to the field of public policy, because we have limited ourselves to economic perspectives, and indeed mostly to one point on that spectrum – the organisation(s) that sponsor(s) the development of systems.

The purpose of **public policy research** is to support the development of policies, and the articulation of programs, that are intended to achieve normative objectives expressed in all of social, economic and environmental terms, and that depend on effective cooperation among disparate organisations, usually across the public, private and often also the voluntary sectors. We already have some research techniques at our disposal that serve such needs, including cost/benefit analysis, risk assessment, and delphi rounds. We need to complement those with further techniques such as technology assessment and privacy impact assessment, which are expressly designed to address the social and environmental as well as the economic dimension, and to cope with the normative as well as the descriptive and explanatory.

# 7  Implications

This section considers the implications of the analysis presented above, firstly for the IS discipline, and secondly for IS practitioners.

## 7.1  Implications for the IS Discipline

This sub-section considers three implications, firstly for the quality of research, secondly for individual researchers, and thirdly for the discipline as a whole and its gatekeepers.

### 7.1.1  Implications for Research Quality

It was noted above that the adoption of the system-sponsor's perspective to the virtual exclusion of other perspectives brings with it a number of important corollaries. The system-sponsor's interests represent objectives. On the other hand, the interests of all other entities that are involved represent, at best, constraints. Other entities that have institutional or market power are regarded as stakeholders, such that their interests are factored in, whereas the interests of other entities are largely ignored.

The perspectives that are excluded, or at least seriously under-emphasised, are most commonly those of users, usees, communities, and the various levels of abstraction along the environment dimension. These interests then surface during the implementation phase as 'impediments' and 'barriers to adoption'. That results in, at best, the costly retro-fitting of features, and 'public education' campaigns, and, at worst, negative return on investment and outright project failures.

The first implication of the analysis of perspectives in IS research is that the entire field of 'impediments' and 'barriers to adoption' is an aberration. Put another way, if the IS discipline's scope and body of knowledge were well-balanced, there would be no such field of study. By expanding the discipline's scope to encompass user, usee, community and environmental perspectives, all relevant parties will be recognised as stakeholders from the outset, and the necessary features will be part of the design, not expensive afterthoughts. There will be of course be mis-analysis, and there will be ineffective designs, and there will be changes in contexts and expectations; but there will be no more late and unpleasant surprises arising fom the absence of the relevant perspectives during the foundation phases of the project.

### 7.1.2 Implications for Individual Researchers

The author contends that, in order to address the deficiencies discussed in this paper, researchers should recognise the following obligations:

(1) **Discover** perspective as an important element of research conception, design, conduct and reporting

(2) **Deliberate** on the alternative perspectives relevant to the context

(3) **Determine** the perspective(s) to be adopted

(4) **Declare** the perspective(s) adopted

Declarations of conflicts of interest are common, but only in relation to funding sources. On the other hand, the AIS Research Code of Conduct at cl.8, on p.8, does not limit the scope to funding sources, but instead says "Declare <u>any</u> material conflict of interest or relationship that might interfere ..." (emphasis added). Objectivity, to the extent that it may be capable of being operationalised, is most usefully treated as being 'explication of bias'. A client relationship is only one of the possible sources of bias. Others include:

- institutional values (business school, social science or engineering faculty, critical theory, ...)
- sub-disciplinary values (MIS, WI, IM, ...)
- school of thought within the discipline
- personal world-view, particularly among economic, social and environmental values

Of course, there are many circumstances in which some aspects are already evident to the reader, e.g. from the author's affiliation, from the explanation of the motivation underlying the research, from the choice of theoretical lens, or from the choice of research method. The nature of the declaration may vary, but the author has the obligation to make clear not only the client, but also the envisaged beneficiary/ies and audience(s) of the work.

### 7.1.3 Implications for the Discipline as a Whole

There are substantial institutional barriers to the broadening of perspectives in IS research. The gatekeepers include editors and program chairs, reviewers, staff selection committees, promotions and review committees, and research funding committees and their assessors.

There are at least a few signs of progress in achieving acceptance. Firstly, there has been very substantial change in conventions over the discipline's 50 years to date. For example:

- at the level of research philosophy, interpretivism, critical theory and design science now complement scientism as acceptable approaches, initially in research-domain-oriented (and hence usually 'C-level') venues, but progressively in A-level venues as well; and

- at the level of research techniques, textual analysis, disciplined literature review / systematic review / meta-triangulation, meta-analysis, case studies and action research have all initially struggled, but have achieved various degrees of acceptance in various levels of venue.

Business journals such as HBR and the Sloan Mngt Review embrace the normative mode and instrumentalist motivations from a business perspective, and have standing within the IS discipline despite that. In my own recent experience, proposing the topic of 'The Missing Perspectives in IS Research' did not result in a Keynote invitation being rescinded. A Panel Session on 'IS Scholars and Policy' is scheduled for ICIS 2015. And searches in the IS literature for exemplars of policy research do not come up entirely empty-handed.

Notably, Agarwal et al. (2012) differs from mainstream IS research not only in the ways identified earlier, but also in that the factors under study were social and political, not economic. They fall specifically within the field of policy studies, not business studies. Yet the Editors of a mainstream IS journal not only had no difficulty in accepting it for publication, but also selected it as the best article of that year. And it was subsequently selected as one of the five AIS Best Information Systems Publications Awards for the year. There are beachheads. But evidence of those beachheads being exploited, through significant numbers of papers that adopt the perspectives of, in particular, users, usees and the environment, through the emergence of specialist venues, through the progressive recognition of those venues, and ultimately in maturation into a recognised instrumentalist approach, is at this stage much more difficult to find.

Senior members of the discipline need to ensure that their natural conservatism and understandable preference for a degree of stability in the discipline's scope are not so strong as to commit IS to ossification. They may be among the slowest to appreciate and declare the perspective that they are adopting in each research project; to consider adopting perspectives other than that of the system sponsor; and to respect and interact with colleagues who adopt alternative perspectives. In their gatekeeper roles, however, senior members of the discipline need to tolerate and respect diversity in the perspectives adopted by other IS researchers, and better still encourage diversity in perspectives.

## 7.2 Implications for Practice

To the extent that alternative perspectives are recognised as being within-scope of the IS discipline, a significant increase in the value of the academic IS literature to IS professionals can be anticipated. The first source of benefits would be an increase in the quality of research through the inclusion of all relevant factors in the design process, leading to smoother implementation, more rapid adoption, and lower system costs.

A further factor would be an increased emphasis on instrumentalist research that has direct relevance to practitioners' needs, and a proportional reduction in other forms of research which have in some cases less direct benefits, and in other cases not a great deal of practical benefit in any case. In addition, IS research would at last make material contributions to the formation and implementation of public policy. The social and environmental dimensions complement the economic, and hence IS would be contributing to holist solutions rather than only serving the needs of individual organisations and strategic partnerships.

# 8  Conclusions

The IS discipline is at serious risk of continuing on an inwardly-oriented spiral towards angel-on-pinhead research, producing too many papers that are of intellectual or methodological interest to a few within the discipline and a few in cognate disciplines, but not to real people. The IS discipline needs to address a whole flotilla of weaknesses.

This paper has examined the notion of the perspective from which a researcher views the phenomena of interest. It has found that researchers generally adopt a single perspective, but do so without making clear what that perspective is, even to themselves, let alone to the readers of the resulting papers. This largely implicit perspective strongly influences the conception of the research, the expression of the research questions, the research design, the analysis of the evidence, and the results.

A pilot assessment was undertaken, to test the author's impression that one perspective dominates IS research. The large majority of a small sample of articles published in AJIS and ACIS in 2014 was strongly oriented towards the interests of the system-sponsor. Alternative perspectives were identified along three dimensions. On the economic dimension, the perspectives of other levels of abstraction are sometimes adopted, particularly those of joint ventures and other forms of strategic partnership. Papers that adopted the perspectives of points on the social dimension, such as users, usees, communities and societies, were uncommon, representing only 4% of the sample. Not one paper in the sample fell on the environment dimension, at any of the possible levels of abstraction such as the biosphere (the presumed ultimate beneficiary of 'green' IT activities) and the troposphere (where climate change is occurring).

Only a single paper in the sample reflected multiple perspectives rather than just one. Examples do exist in the literature, however, including a recent prize-winner. This represents a potential breach in the wall of research driven by the interests of system-sponsors.

Research approaches that extend beyond the descriptive, explanatory and predictive were reviewed. The constructivist technique of design science has to date mostly been tied to the system-sponsor perspective, whereas action research has an association with interpretivism and inherently reflects multiple perspectives. The broader field of problem-driven, instrumentalist research was depicted as a step beyond conventional interpretivist approaches, in that it not only recognises the existence of multiple interests, but also begins with a problem and seeks solutions that achieves balance among the interests. Critical theory research was noted as being explicitly concerned with power-structures and with altering pre-existing balances. The progressive extension of IS research into normative mode was then combined with the recognition of the social and environmental dimensions. This led to the conclusion that public policy research is a natural part of the scope of the IS discipline.

The paper concludes that the dominance of the system-sponsor perspective has resulted in lower quality than could otherwise be achieved, and is causing the discipline to miss opportunities for contributions even on the economic dimension, but particularly in the social and environmental areas. Members of the IS discipline need to discover perspective as an important element of research conduct, deliberate on the alternative perspectives, determine the perspective(s) to be adopted, and declare the perspective(s) as part of their reports on the research. Senior members of the discipline need to ensure that they do not confuse their role as gatekeeper with the notion of palace guard, and that they permit entry to contributions that satisfy quality tests appropriate to the subject-matter.

## Acknowledgements

Valuable feedback on an early draft of this paper was received from several colleagues, notably Robert Davison, Frank Land and Doug Vogel.  The emergent thesis and its many remaining challenges remain the responsibility of the author.

# Appendix 1: Research Questions in AJIS Vol. 18 (2014) – 38 Papers

**Papers on Research Methods, the IS Discipline, Teaching Cases (2)**

**Theory Papers – System-Sponsor Perspective Predominant (3)**

**Empirical Papers – System-Sponsor Perspective Dominant**

- **Studying Organisational Phenomena (22)**

- **Studying Human Phenomena (2):**

- What factors affect customers' continued use of Internet Banking?

- Does alignment of sensing and responding result in greater customer satisfaction?

**Empirical Papers – Some Other-Party Perspective Also Apparent (7):**

- Are we evaluating Australian interventions for women in computing?  (Some reflection of the employee perspective)

- What are the education needs of the business analyst?  (Some reflection of the employee perspective)

- To what extent do particular factors influence university staff adoption of iPads in their work?  (Some reflection of the employee perspective)

- How do you measure manager satisfaction in using the enterprise resource planning (ERP) system?  (Some reflection of the employee perspective)

- What factors influence Green IT assimilation?  (Some reflection of the environmental perspective)

- Do ICT intervention programs have an influence on ICT career intentions?  (Some reflection of the student perspective)

- How are our students using mobile technologies for teaching and learning?  (Some reflection of the student perspective)

**Social Dimension (2):**

- What measures may protect privacy of the users in the context of RFID?

- What are the potentials of assistive technologies for seniors' independent living?

**Environmental Dimension (0)**

## Appendix 2:    Research Questions at ACIS (2014) – 36 Papers

**Papers on Research Methods, the IS Discipline, Teaching Cases (1)**

**Theory Papers – System-Sponsor Perspective Predominant (10)**

**Empirical Papers – System-Sponsor Perspective Dominant**

- **Studying Organisational Phenomena (9)**

- **Studying Human Phenomena (9):**

- What information classification scheme will help SNS designers?

- How does elearning compare to a face-to-face setting in enabling students to learn the logic of Business Process Management?

- What factors influence students' perceptions of IT and IT Careers?

- Who uses the Community-Based Environmental Monitoring system, how, and how might it be improved?

- Why are recordkeeping systems experiencing different rates of acceptance and utilization by end users?

- Can we improve participation in university course surveys using mobile tools?

- What effect do different levels of user representations have on users' sharing behaviour on a C2C platform?

- What are the key factors in the adoption of wireless handheld devices in healthcare?

- What are the key factors in User Adoption of Call-taxi App?

**Empirical Papers – Some Other-Party Perspective Also Apparent (6):**

- How are Electronic Patient Records used during nursing handover?  (Some reflection of the professional employee perspective)

- What concepts are discussed on Healthcare Social Question Answering services?  (Some reflection of the user perspective)

- What role does social media play in an election?  (Some reflection of the electors' perspective)

- What use do senior citizens make of social media and how does it affect their life satisfaction?

- How can we enhance information flow within the agriculture domain?

- What are the motivations for travel experience sharing behaviour on social media?  (Some reflection of the user perspective)

**Social Dimension (1):**

- What mental eHealth services will best meet the future needs of regions?

**Environmental Dimension (0)**

## Copyright